# 'Pathway drug cocktail': targeting Ras signaling based on structural pathways

Ruth Nussinov[1,2,*], Chung-Jung Tsai[1] and Carla Mattos[3]

[1]Basic Research Program, SAIC-Frederick, Inc. Cancer and Inflammation Program Center for Cancer Research, National Cancer Institute, Frederick, MD 21702

[2]Sackler Inst. of Molecular Medicine Department of Human Genetics, Sackler School of Medicine, Tel Aviv University, Tel Aviv 69978, Israel

[3]Department of Chemistry and Chemical Biology, Northeastern University, Boston, Massachusetts, 02115

**Abstract**

Tumors bearing Ras mutations are notoriously difficult to treat. Drug combinations targeting the Ras protein or its pathway have also not met with success. 'Pathway drug cocktails', which are combinations aiming at parallel pathways, appear more promising; however, to be usefully exploited, a repertoire of classified pathway combinations is desirable. This challenge would be facilitated by the availability of the structural network of signaling pathways. When integrated with functional and systems-level clinical data they can be powerful in advancing novel therapeutic platforms. Based on structural knowledge, drug cocktails may tear into multiple cellular processes that drive tumorigenesis, and help in deciphering the interrelationship between Ras mutations and the rewired Ras network. The pathway drug cocktail paradigm can be applied to other signaling protein targets.

***Correspondence**: Ruth Nussinov, National Cancer Institute, Frederick, MD, 21702; E-mail: NussinoR@helix.nih.gov; Fax: 301-846-5598; Phone: 301-846-5579

**Bridging cellular, structural and dynamic data at multiple scales**

Ras proteins are essential components of signaling networks controlling cellular proliferation, differentiation, and survival (Box 1). Deregulation of Ras-dependent signaling is common in tumorigenesis [1], and to date, tumors bearing Ras mutations have been among the most difficult to treat [2]. Efforts focusing on single pathways downstream of Ras have been only partially successful, as can be seen with drugs, such as PLX-4032, that target Raf kinase, which binds directly to Ras in normal cells: while demonstrating some benefit in the treatment of melanoma caused by BRAF mutations these compounds have not been successful for Ras mutant tumors [2]. This example emphasizes the challenge: turning off Raf kinase, a major downstream effector of Ras action and thus a drug target, can activate the mitogen-activated protein kinase (MAPK) pathway in some Ras mutants, bypassing the drug action. Further, MAP kinase kinase (MEK) inhibitors can block the Ras-MAPK pathway; however, they can activate the phosphoinositide 3- (PI3) kinase via epidermal growth factor receptor (EGFR) and other receptor tyrosine kinases by uncoupling a negative extracellular signal-regulated kinase (ERK) feedback loop. Such drug resistance mutations and feedback loops examples point to the need for not one but multiple combinations of drug therapies. Nonetheless, which pathways to combine (to avoid alternative routes by drug resistant mutations) and which protein targets to select within these (to minimally perturb normal functions) present a daunting challenge. A classified collection can be a first step toward a 'pathway drug cocktail'. A repertoire of pathway combinations, coupled with minimally toxic protein targets within them, is a formidable mission, underscoring the urgency for transformative strategies.

Here, we point to the merits of an integrated, structure-based strategy that (i) will create a more complete structural map of the Ras pathways and their cross-talk; (ii) will map the mechanisms of how the distinct oncogenic Ras mutations work: directly in the active sites or at post-translational modification (PTM) regions, or indirectly, by allosterically trapping Ras in a constitutively activated state; and (iii) will target scaffolding proteins, such as kinase suppressor of Ras (KSR), a positive modulator of the Ras/ERK signaling. Scaffolding proteins, which allosterically regulate enzymatic' activities [3], are often overlooked in pathway maps, as in the case of KSR (Figure 1), and in drug development. When integrated with biochemical and cellular data, and related to oncological expression, structural pathways can help fill in missing protein components [4] (Figure 2), clarify regulation [5], predict possible side effects, and uncover combination targets. Combined with biochemical and biophysical data, structural pathways can help in bridging mechanistic and dynamic gaps, implementing powerful technologies toward crucial translational applications.

**Ras: conformation and allosteric mechanism**

The structural features of Ras and the mechanisms through which it is regulated by GEFs and GAPs have been worked out at level of the interacting domains (Figure 3) [6, 7]. The switch associated with the nucleotide-bound state (GTP→ON; GDP→OFF), unifies GTPases across cellular functions, including signal transduction, cell motility, trafficking, and protein biosynthesis [8], although the mechanistic details of how GEFs and GAPs regulate the Ras family GTPases vary [9]. Current data are based primarily on a single static model for each protein or complex, overlooking conformational dynamics and allosteric effects which may bias pathway selection [3].

The Ras catalytic domain is composed of the canonical G-domain structure with 6 β-strands flanked by 5 α-helices and 10 connecting loops [10]. The P-loop switch I (L2) and switch II (α2 and L4) contain the active site for GTP hydrolysis and interaction sites for effector proteins, including Raf and PI3K (Figure 4a). These structural elements are within the first 'effector lobe' half of the catalytic domain, which is 100% conserved across Ras isoforms [11, 12]. On the other side of the molecule, fully within the second, 90% identical 'allosteric lobe' half of the catalytic domain, is an allosteric site [13] and several affinity hot spots that interact directly with membrane components [11, 14]. The membrane interacting hot spots contain many residue differences between the Ras isoforms outside of the hypervariable C-terminal region (Figure 4a).

The interplay of interactions between the two lobes of the catalytic domain links the active site for GTP hydrolysis with membrane interacting sites on Ras [13]. The switch I and switch II regions are intrinsically disordered [15, 16], resulting in a disordered active site that accounts for the very slow rate of intrinsic hydrolysis measured *in vitro* [17]. This has for years justified the prevailing assumption that intrinsic hydrolysis in Ras is biologically irrelevant, and obscured subtle conformational changes that may be pathway specific and affected differently by oncogenic mutants [18, 19]. The allosteric switch, with helix α3 and loop L7 connecting the allosteric site at the membrane interface to switch II, communicates between the cell membrane and the active site and appears to be specific to the Ras/Raf/MEK/ERK pathway [13, 18]. Raf binds to Ras at switch I, but not switch II, with nanomolar affinity [20]. This affinity is about 1000-fold greater than the affinity of Ras for GAPs [21], which in order to promote hydrolysis of GTP would have to displace Raf to occupy an overlapping binding site. While currently still unproven, the working model is that Raf itself, in concert with a membrane component interacting at the allosteric site, promotes the intrinsic hydrolysis of GTP in the absence of GAP [13, 18, 19]. Raf stabilizes the conformation of switch I associated with intrinsic hydrolysis of GTP [22], but switch II, containing the catalytic residue Q61, remains disordered. The Raf-bound Ras propagates the signal in the presence of a disordered switch II with an impaired active site, until binding of $Ca^{2+}$ and an unknown membrane component to the allosteric site promotes ordering of switch II, enhancing GTP hydrolysis and signal termination. In this model of signaling through the Ras/Raf/MEK/ERK pathway, Raf plays a critical role in stabilizing a switch I conformation important for the activation of GTP hydrolysis by the allosteric switch mechanism. It thus behaves as a scaffold in the broad sense of a protein that increases the population of certain conformational states [3].

**How mutations activate Ras even in the absence of RTK signals**

The C-terminal hypervariable region (HVR) guides localization of the isoforms in distinct membrane compartments. The allosteric switch mediates between the membrane anchor and the effector lobe. It also may link oncogenic mutations in the isoforms and their cancer-specific signaling pathways [23]. Box 2 provides an overview of Ras somatic mutations in cancer. Mutants may promote constitutive activity by impairing GTP hydrolysis in the presence or absence of GAPs via distinct mechanisms, resulting in different pathway activation. G12V and Q61L mutants have distinct effects on the Ras/Raf/MEK/ERK pathways in NIH3T3 cells: whereas RasQ61L results in saturated ERK phosphorylation, RasG12V increases p-ERK levels but to a much lesser extent [18]. In the presence of Raf the switch II in RasQ61L adopts an anti-catalytic conformation with L61 at the center of a hydrophobic core that closes over the nucleotide, isolating it from bulk solvent [19]. In contrast to the distinct behavior of the two mutants with respect to the Ras/Raf/MEK/ERK pathway, they have similar effects in signaling through the Ras/PI3K/Akt pathway [18]. Although oncogenic mutations in residues other than G12, G13 and Q61 are rare (Box 2), several human tumors contain mutations in K-Ras or N-Ras switch II, helix 3 and the allosteric site [23]. Prominent examples are the two arginines depicted in Figure 4b. Human tumors with R68S in K-Ras and R68T in N-Ras, and R97G in N-Ras [24] presumably are unable to promote ordering of the active site in Raf-bound Ras, leading to constitutive signaling through the Ras/Raf/MEK/ERK pathway. Effectors such as PI3K and RalGDS, bind to Ras through both switch I and switch II with affinities in the micromolar range, comparable to that of the Ras/GAP complex [21, 25, 26]. GAP can compete with these effectors to promote hydrolysis of GTP and turn off signaling. Thus, oncogenic mutations may affect pathways differently. Structural pathways may provide sets of pathway drug cocktails targets for such isoform-specific cancer mutations.

**Targeting Ras: failures and promise**

Ras as a target for drug discovery has commanded intensive focus in the last 20 years with very little success [27]. Ras proteins undergo isoform-specific lipophilic post-translational modifications, including irreversible farnesylation and reversible palmitoylation, [28, 29]. However, attempts to target the Ras family by focusing on farnesyltransferase inhibitors have not turned out as well as hoped: peptidomimetics experienced problems with cell permeability, or were rapidly degraded in the cell; cross-prenylation took place which resulted in a persistent membrane localization of K-Ras and N-Ras along with upregulation of downstream signaling. Therapeutic approaches targeting prenylation and post-prenylation modifications as alternatives [30] encountered different complications: the higher frequency of prenylation in membrane-anchoring proteins necessitates higher active antitumor concentrations which lead to toxic effects [2]. Inhibition of some postprenylation processing enzymes has had only modest effects; of others, like isoprenylcysteine carboxyl methyltransferase, led to more encouraging effects, though still overall limited success. Targeting the regulation of GDP/GTP exchange did not meet with success either. These largely failed attempts to interfere with constitutive activation have led to greater focus on pathways downstream of Ras; however, also achieving only limited effectiveness. Blocking single pathways faces cell proliferation: network degeneracy allows resistance-driven pathway interchangeability to take over, establishing cancer robustness. Collectively, pharmacokinetics, pharmacology, toxicity, and the persistence of drug resistant mutations, argue for multiple novel strategies, including 'pathway drug cocktails'; i.e. targeting combinations of interconnected pathways. An encouraging example relates to the combined inhibition of the Raf-MEK-ERK and the PI3' kinase signaling pathways, which evades their feedback loop crosstalk. This was the case not only for activated receptor tyrosine kinase, but also for oncogenic Ras mutations [31]. Even though the superiority of dual pathway inhibition has been obvious, and its clinical prospects considered beneficial, to date a broad, comprehensive and integrative approach has yet to be mapped out. This should be possible with complete structural pathways.

With the poor success of the strategies described above, and in light of the recent progress in understanding the dynamical details associated with conformational states in Ras, it may be time to redouble efforts to directly target this important signaling hub protein (Box 1). Part of the problem to date may be that most of the efforts are geared toward obtaining drugs that bind in the active site region [32], which is disordered in solution [15, 16]. The majority of the hot spots are either at the interface between the effector and allosteric lobes of Ras or entirely in the allosteric lobe [11]. Perhaps targeting those sites of protein membrane interactions important for the active state of Ras, or developing molecules that selectively stabilize the catalytically competent state of the allosteric switch to alleviate the constitutively activated state associated with the RasQ61L mutant in the presence of Raf, would meet with greater success. Recent work has shown that it is possible to modulate the state of the allosteric switch with small molecules in the crystal [33], suggesting targeting a specific signal transduction pathway within a comprehensive pathway drug cocktail strategy.

**Scaffolding proteins in the Ras network as targets**

Scaffolding proteins are typically found in multienzyme complexes in signaling pathways. They work via a conformational biasing mechanism [3]. They do not have (primary) active sites; however, they enhance catalysis via population shift and by creating a favorable enzymatic environment [34]. KSR is a scaffolding protein that stimulates Raf phosphorylation of MEK and coordinates the Raf/MEK/ERK assembly. In the KSR2(KD)/MEK1 heterotetramer [35], KSR2 is inactive, burying the MEK activation segment. The binding of B-Raf, a Raf isoform frequently mutated in cancers, to KSR leads to conformational change that releases and exposes MEK's activation segment, allowing it to be phosphorylated by another catalytic Raf. KSR is important for the increased MEK1 phosphorylation as shown by adding the ASC24 drug to KSR(KD)/MEK1. ASC24 disrupts the KSR2(KD) homodimer interface thereby inducing a conformational change which allows increased phosphorylation of MEK1 by kinase-impaired B-RafK483S. A Ras inhibitor that would act like KSR(R718H) mutation could be an antagonist of the B-Raf-induced allosteric switch [35]. Thus, even though often not included in 'classical' pathways, scaffolding proteins can be drug targets and the more complete structural pathways would have them.

**Structural networks and acquired resistance mechanisms**

Acquired drug resistance can take place via two distinct mechanisms: genetic and nongenetic [36]. The genetic mechanism takes place through somatic alterations associated with resistance to targeted therapies; the nongenetic mechanisms are driven by changes such as in epigenetics, RNA splicing, metabolism, and protein post-translational modifications. Although the causes differ, both mechanisms involve a rewiring of the network. The Ras/Raf/MAPK pathway provides examples for both mechanisms. For the genetic mechanism, resistance associated with B-Raf inhibitors can be MAPK pathway-dependent or independent [37]. In the MAPK pathway-dependence mechanism, B-Raf or MEK inhibitor concentrations that normally suppress cell growth can reactivate p-ERK. Mutation of the gatekeeper residue of B-Raf, or those that abolish dimerization or cause aberrant B-Raf mRNA splicing can disable B-Raf's interaction with the inhibitor and cause resistance [38]. Other examples of MAPK-dependent mechanisms of B-Raf inhibitor resistance involve acquired mutations in N-Ras [39], or mutations in MEK that increase catalytic activity [40]. ERK-independent mechanisms of resistance to B-Raf inhibitors typically involve parallel signaling. Pathway involvement is also seen in amplified oncogene K-Ras that can confer resistance to MEK inhibitors by signaling, rewiring through ERK1/2 [41, 42]. Nongenetic mechanisms of acquired drug resistance can involve switching dependencies from one activated kinase to another via an addiction mechanism, or can depend on more than one convergent pathway. Unlike addiction switching, which often takes place only in a small percentage of the tumor cells, pathway crosstalk occurs in the majority of tumor cells during treatment by anticancer drugs.

**The use of drugs in combination for the treatment of cancer is not new**

The use of drugs in combination for the treatment of cancer has a long history [43]. The National Cancer Institute has published a summary of drug combinations, as well as common combinations for colon and rectal cancer. To find the more effective treatments for cancer, 5,000 combinations of 100 existing cancer drugs have been tested. Recently, using high-throughput screening techniques, 150 drugs that were genotype-selective for Ras or B-Raf mutations were searched for pairs that could inhibit metastasized melanoma. Surprisingly, including statins in the drug combinations killed Ras-driven melanoma [44], albeit at concentrations ten times higher than are considered safe in humans. In another study, combination of allosteric MEK, and B-Raf inhibitors that target the same pathway was found to be effective [45]. Sequential, rather than simultaneous, application of combination drugs can also lead to better response due to the prior cell sensitization [46].

While such comprehensive strategies can obtain beneficial results, a strategy based on more complete information of the cellular network may be expected to allow for more accurate and deliberate targeting of specific cancers. Developing new drugs based on the pathways, or mapping the drugs that are already available onto the pathways and picking potentially promising combinations, may shorten discovery timescales.

**Constructing structural pathways and choosing combinations**

The construction of structural networks of signaling pathways is not only a concept. Some methods are already in place [47, 48]. Although the computational time involved in docking of protein complexes on a large scale are prohibitive, other knowledge-based strategies, such as PRISM (PRotein Interactions by Structural Matching), have been developed and already showed their merit in constructing several signaling pathways, including apoptosis [4] and E2-E3 interactions. The results indicated that they provided more complete information, obtained interactions missing in the KEGG pathway, and led to new insight (Figure 2). Such methods are based on the principle that structural motifs recur in nature, in protein cores and in protein-protein interfaces. Known interfaces from the Protein Data Bank (PDB) can thus serve as templates, against which surfaces of candidate interactions of proteins experimentally known or suspected to play a role in the pathway, can be compared. The predictions are followed by refinement and energy evaluation, and validated by protein interaction databases and the literature. Such a procedure can be carried out if experimental or high quality modeled structures of the proteins are available. Key caveats in large scale knowledge-based approaches are that they are only able to handle a limited extent of protein flexibility, and the motifs should be present in the PDB. Graph theoretic techniques can next find the shortest paths between two vertices (nodes). Combinations of these can provide parallel pathways. Although from the computational standpoint this problem is tractable, drugs developed based solely on compilation of combinations of the pathways are unlikely to succeed. Below we provide a few examples illustrating this point.

*Compensatory effect of oligomeric state*s. B-RafV600E activation is Ras independent, driving tumors by deregulating ERK signaling. Raf inhibitors potently inhibit RafV600E monomers and ERK signaling, but not dimers, which are still active in these tumors. These lead to a rebound in ERK activity by relieving a feedback loop, which culminates in a new steady state, where ERK signaling is elevated compared to its low level after the initial Raf inhibition. In this state, ERK signaling is Raf inhibitor resistant, springing back Ras activity, and making ERK signaling ligand-dependent. Combined inhibition results in enhancement of ERK pathway inhibition and antitumor activity [49]. On their own, pathways (e.g., KEGG, Figure 1, for the Raf/MEK/ERK pathway) and graph theoretic strategies will overlook the compensatory effects of dimers when monomers are inhibited; however, such unexpected Raf dimer formation might be straightforwardly discovered from structural pathway prediction. Dimer prediction would be similar to the prediction of XIAP inhibitor binding to Caspase-7 and -3 through the same baculoviral IAP repeat-containing protein 2 (BIR2) surface (thus not taking place simultaneously), which is not the case for X-linked inhibitor of apoptosis protein (XIAP) binding to Caspase-9 and -3, or -7; or Caspase-8, which interacts with BH3 interacting-domain death agonist (Bid) and Procaspase-3 via different Caspase-8 surfaces (Figure 2, right hand-side structures).

*Relief of feedback inhibition.* Raf inhibitors, which demonstrate elevated specificity for B-RafV600E can have an opposite effect in cancer cells with wild-type B-Raf, including those with oncogenic Ras mutations [2, 49-51]. The rebounding of Raf/MEK/ERK signaling in Ras mutant cancer cells by Raf inhibitors precludes these inhibitors in the treatment of cancers in which Ras is mutated. Treating melanoma tumors harboring B-RafV600E mutations with the pan-Raf inhibitor PLX-4032 resulted in multiple resistance mechanisms [52], including enhanced receptor tyrosine kinase signaling and mutational activation of N-Ras [39]. Relief of feedback inhibition will be overlooked by combination strategies. Structural pathway prediction may be able to discover additional new relief of feedback (or positive feedback) to alter the core processes, as in the case of apoptosis (Figure 2, IAP, bottom, right).

*Mechanism of activation.* Broadly, there are four possible mechanisms for kinase activation: (i) graded, or incremental through several events, such as dephosphorylation of a certain region, phosphorylation of the activation loop, binding of a ligand, etc. Each event leads to higher activity; (ii) an "OR-gate", where the kinase is activated in an all-or-none fashion after either phosphorylation or binding of phosphorylated ITAM peptides; (iii) activation follows an "AND-gate" mechanism. Both events are needed for activation: loop phosphorylation and proper binding [53]; (iv) through interaction with the membrane. Src kinase activation takes place as a graded switch; Syk kinase as an "OR-gate" switch; Tec kinase activation as an "AND-gate" switch; Raf kinase activation via interaction with the membrane. The mechanism of activation affects the downstream signaling pathway. It relies on structural detail of the interaction that a structural pathway can provide; but which is unavailable in conventional pathways. This can be seen by shared versus non-shared binding surfaces (Figure 2, right hand-side), modeled in the structural pathways.

*Allosteric vs orthosteric (full) activation.* Currently, in the absence of structural information it is unclear how Ras-GTP activates Raf. In principle, kinase activation can take place via two mechanisms. The first is orthosteric, through autophosphorylation; once Ras activates Raf, the necessity for Ras is abrogated. Phosphorylation of key residues located in the activation loop would organize and stabilize the active conformation. The second, more likely mechanism is allosteric. Because the perturbation caused by Ras binding to the Ras-binding domain of Raf (RBD) propagates through the flexible linker to the catalytic domain, it is expected to result in a conformational change. Indeed, even orthosteric autophosphorylation probably follows allosteric activation of Raf through Ras binding to RBD. Structural information on the conformational consequences of Ras binding to Raf would resolve this question: if Ras binding to Raf makes the αC-helix move, it suggests an allosteric mechanism. Under such circumstances, Ras mutations that prevent catalysis of GTP hydrolysis by GAP abolish Ras regulation of Raf activation, making it constitutive. This highlights the importance of complete structural data, as mentioned above for the interaction surfaces in the apoptosis pathway. Structural information would also help to understand the pathway of the different isoforms.

The criteria for drug combinations have been thoroughly reviewed, including those for the same protein, same pathway, and different - related and unrelated - pathways, and drug-mediated molecular interaction profiles [54]. These include drug types, pharmacodynamics synergism, complementary drug actions, additivity and agonistic actions, and more, including pathway analysis, all documented with examples. Structural pathways can provide additional data, which may relate to protein oligomeric states; knowledge of whether partner proteins share a binding site, and thus cannot take place simultaneously in an assembly as well as multiple sites; allosteric effects; mutant forms and their modes of activation and the presence of additional protein targets in the pathways, such as scaffolding proteins.

**Concluding remarks**

The ultimate aim of cracking the Ras mutations code is challenging. The ability to predict the relationship between specific mutational spectrum and alternate signaling pathways is still a faraway wish. The reward of such a translational code would be an embodiment of predictive personalized Ras therapeutic repertoire. Inroads into this problem can only be obtained with the help of structural knowledge. At the molecular level, resistance mutations work by leading to a conformational change that can shift the conformational ensembles [55] of Ras and thus its preferred binding partners, or pattern of post-translational modifications, which in turn play key roles in localization and network rewiring. Although most mutations do not significantly affect the conformational behavior, oncogenic and drug resistant mutations do.

Approximately 20% of all human cancers have oncogenic mutations in the RAS gene. To date, therapeutic approaches based on the inhibition of Ras-mediated signaling, including Ras processing, and its post-translational modification enzymes and downstream Ras effectors have not been productive [56]. Two major factors are responsible: toxicity of drugs which when given in effective dosage, not only hit their intended target, but also disable normal functions, and drug resistant mutations that circumvent drug action. The cross-connected cellular network has extensive redundancy of signaling pathways also in healthy cells; however, in cancer these alternative pathways take over, losing control. To combat pathway redundancy drug combinations strategies are becoming common. Combinations include orthosteric and allosteric drugs for the same [57, 58] or for different proteins, in the same, or in different pathways [59]. Here, we suggest compiling a repertoire of pathway combinations based on the construction of the cellular structural network, integrated with functional/biochemical data. Much of the structural data is already available in the PDB; but we, as a community of structural biologists, can focus greater efforts in filling in specific gaps for targeted pathways such as those involving Ras. The structural and functional genomics initiatives have led to a fast increase in the number of experimental structures at high resolution. High quality homology modeling enhanced by low-resolution structural constraints of proteins and assemblies complements and enhances the structural universe.

Concerted efforts putting these together would obtain the cellular structural universe; coupled with conformational behavior of oncogenic mutants, it may provide the much-needed insight into the distinct expression and rewiring profiles relating mutants to cancer types. This may help forecast which effector pathways are activated, and how intensively [60]. Yet, it behooves us to recall that structural networks are not sufficient; the fact that resistance can take place via feedback loops underscores the paramount requisite for complementation by functional data. Collectively, employing a range of techniques over multiresolution scales, cellular processes that drive tumorigenesis can be better understood and targeted by 'pathway drug cocktails'.

## Acknowledgements

Thanks to Christian W. Johnson for making Figures 3 and 4. This project has been funded in whole or in part with Federal funds from the National Cancer Institute, National Institutes of Health, under contract number HHSN261200800001E. The content of this publication does not necessarily reflect the views or policies of the Department of Health and Human Services, nor does mention of trade names, commercial products, or organizations imply endorsement by the U.S. Government. This research was supported (in part) by the Intramural Research Program of the NIH, National Cancer Institute, Center for Cancer Research (Ruth Nussinov) and by the extramural program under grant number R56 CA096867 (Carla Mattos).

**Glossary**

EGFR – Epidermal growth factor receptor; MAPK - mitogen-activated protein kinase; ERK- Extracellular Signal-Regulated Kinases; MEK - MAPK/ERK kinase; PTM – Post-translational modification; KSR – Kinase suppressor of Ras; KEGG – Kyoto Encyclopedia of genes and genomes; GAP - GTPase Activating Protein; GEF - Guanine-Nucleotide Exchange Factor; SOS - Son of Sevenless; HVR – C-terminal hypervariable region; RalGDS - Ral guanine nucleotide dissociation stimulator; PRISM – Protein Interactions by Structural Matching; IAP – Inhibitor of Apoptosis; PDB – Protein Data Bank; TNF – tumor necrosis factor; MEKK/SEK/JNK - Jun N-terminal Kinases, PI3K -Phosphatidylinositol 3-Kinase; NF-KappaB - Nuclear Factor-Kappa B, Raf/MEKK1/IKK -I-KappaB Kinase; GRB2 - Growth Factor Receptor-Bound Protein-2.

**Box 1: Ras as a hub protein**

Ras family proteins are among the most important hubs in cellular signaling pathways. RAS genes encode monomeric GTPases that function as membrane-associated molecular switches [61, 62]. Upstream Ras signaling initiates upon activation of cell surface receptors, such as RTK and TCR, and cycles between ON, GTP-bound and OFF, GDP-bound states. Interactions of GTP-bound Ras with effector proteins may lead to cell proliferation. GTP→GDP hydrolysis leads to the OFF state. Ras activates key signaling pathways, including Raf/MEK/ERK, MEKK/SEK/JNK, PI3K /Akt/NF-KappaB pathway, p120-GAP/p190-B/Rac/NF-KappaB, and Raf/MEKK1/IKK/I-KappaB/NF-KappaB. Most growth factors that signal through RTKs or heterotrimeric GPCR stimulate Ras by recruiting the GEF SOS to the membrane. SOS exists in a complex with the scaffolding protein GRB2. Upon receptor activation, the GRB2/SOS complex is translocated to the membrane by binding of GRB2 to tyrosyl-phosphorylated residues in RTKs or other scaffolding proteins. GTP-bound Ras recruits and activates Raf. Raf initiates a cascade of phosphorylation by phosphorylating MEKs, which phosphorylate ERKs. Phosphorylated ERK moves from the cytoplasm into the nucleus phosphorylating transcription factors, including Elk1, ATF-2, FOXO4, NFAT, STAT3 and C-Jun [63], which turn on transcription of specific sets of genes. The non-structural Ras interactome has been compiled, including proteins that regulate Ras GDP/GTP cycling, catalyze its post-translational modification or serve as immediate downstream effectors [63, 64]. Among the key downstream effector signaling, is the Raf serine/threonine kinase, which binds to activated Ras, turning on the highly conserved Raf/MEK/ERK cascade; the PI3' kinase/PIP3 and the Tiam, Akt and PDK1 substrates-cascades; Tiam1, Ral guanine nucleotide dissociation stimulator (RalGDS), and PLCε. All interact directly with Ras [2, 63]. To date, pathways [65] are incomplete: proteins are missing, and structural data largely relate to monomers and small complexes (Figure 3). Conformational flexibility that may be associated with allosteric regulation is also lacking. The PDB abounds with crystal structures of Ras: over 50 H-Ras, and the recently solved K- and N-Ras. These confirm the high similarity between the isoforms [2, 66], supporting the notion that the functional differences are driven by the C-terminal HVR, which contains the linker region and the membrane-interacting lipid anchor.

**Box 2: Somatic mutations in cancer**

The Ras protein family includes H-Ras, N-Ras, and splice variants, K-Ras4A and K-Ras4B [67]. K-Ras mutations are the most prevalent (21%), N-RAS (8%) and H-RAS the least (3%) [23, 63, 68]. Mutated isoforms tend to associate with particular tumor types: K-Ras mutations are in the majority of pancreatic ductal adenocarcinoma and with significantly high percentages of lung and colon tumors; however, they are rare in bladder tumors, which is where mutated H-Ras is the most frequent. N-Ras mutations are often observed in hematopoietic tumors and in malignant melanomas, whereas the rate of K-Ras or H-Ras mutations in these tumors is marginal [1].

Oncogenic mutations [1, 23] are concentrated around two hotspot regions (codons 12 and 61). In K-Ras, G12-G13 accounts for ~99% of the mutations, whereas those at Q61 only 1% [24]. Other mutations also exist, including L19F and T20A combination in colorectal carcinoma [69]. In N-Ras Q61 is at 60%, G12 at 24% and G13 at 13% [24]; in H-Ras G12 (54%), Q61 (34.5%) and G13 (9%). Overall, these mutations implicate poor prognosis. However, exon 4 mutations in K-*ras* have a more favorable prognosis [1, 70]. The different consequences suggest pathway variability. However, how pathway dominance shifts is still unclear.

**Figure Legends**

Figure 1. Classic-style illustration of the Ras/Raf/MEK/ERK signal transduction pathway, redrawn from a KEGG diagram. The pathway is incomplete: it misses proteins, such as KSR, and is devoid of interaction details. The usefulness of such a diagram can be contrasted with that depicted in Figure 2 which provides the structural pathways, where the structures of the proteins are available. Ras is highlighted. Abbreviations: NGF: nerve growth factor; BDNF: brain-derived neurotrophic factor; NT3/4: neurotropin 3/4; EGF: epidermal growth factor; FGF: fibroblast growth factor; PDGF: platelet-derived growth factor; CACN: calcium channel subunits family; TrkA/B: Tropomyosin receptor kinase A/B; EGFR: epidermal growth factor; FGFR:

fibroblast growth factor; PDGFR: platelet-derived growth factor; GRB2: Growth factor receptor-bound protein 2; G12: G-protein G12 subfamily; RasGRF: RAS Guanine-nucleotide exchange factors; RasGRP: RAS guanyl nucleotide-releasing protein; SOS: Son of Sevenless; Gap1m: GTPase activating protein 1m; GNGEF: Guanine nucleotide exchange factor; PKC: protein kinase C; NF1: neurofibromin Type 1; pl20GAP: pl20-GTPase activating protein; DAG: diacylglycerol; Rap1: Ras-proximate protein 1; Mos: Proto-oncogene serine/threonine-protein kinase mos; MP1: MEK partner 1; ERK: Extracellular signal-regulated kinases; PTP: protein tyrosin phosphatase; MKP: mitogen-activated protein kinase phosphatase; STMN1: Stathmin 1/oncoprotein 18; cPLA2: phospholipase A2; MNK1Q: mitogen-activated kinase 1; RSK2: ribosomal s6 kinase 2; CREB: cAMP response element-binding protein; Elk-1: extracellular-regulated kinase; Sap1s: serum response factor-associated protein-1 (or ELK4); SRF: serum response factor.

Figure 2. The structural pathway of apoptosis constructed using PRISM [4]. Initially, the KEGG pathway diagram is used. In the redrawn diagram, the proteins are in blue ellipsoids, connected by black links. Where structures are available in the PDB, PRISM [47] is able to predict the structure of the complex, based on template interfaces available in the PDB. These interactions are shown in green. The interactions depicted by pink lines are not in KEGG, but are predicted by PRISM for proteins known to be in the apoptosis pathway. Subsequent searches in the literature verified these interactions. The functional roles of the interactions of the Inhibitors of Apoptosis (IAP, orange ellipsoids), a family of functionally and structurally related proteins which serve as endogenous inhibitors of programmed cell death, are as follows: the interaction between IAP and Akt/PKB and IAP-IKK implies action by enhancing phosphorylation, thus activation of NFκB, leading to pro-survival signal; edges ending by a bar, imply that binding of IAP directly interferes with binding e.g. of TNFα to TNF-R1; of CASP-8 to TRAF2, CASP-7, CASP-9; of CASP-3 to CASP-6; etc. (CASP-3, -6, -7 are activated by CASP-8 or -9). The adaptor proteins are in the Death-Inducing Signaling Complex (DISC). DISC activates CASP-8. The predicted FLIP-Procaspase-8 complex was not in KEGG. Inhibition of IAP would promote apoptosis via a number of pathways, and IAP is an important drug target [71]. As can be seen in the map, IAP is an actor in a positive feedback loop. The structures of the complexes of proteins highlighted by faded background colors are depicted on the right. On the top right, the prediction of the XIAP inhibitor binding to Caspase-7 and -3 is shown to be through the same BIR2 surface (thus not taking place simultaneously). This is not the case for XIAP binding to Caspase-9 and -3, or -7 below, where the interaction takes place via different XIAP surfaces; or Caspase-8, which interacts with Bid and Procaspase-3 via different Caspase-8 surfaces, at the bottom right. Prism is a knowledge-based strategy, developed based on the observation that similar to single chain proteins, protein-protein interfaces also present a set of recurring motifs. Cn stands for calcineurin. An additional prediction not shown here is between IAP and TNFα. Further details on the construction of the apoptosis pathway are given in Acuner Ozbabacan et al. [4]. Abbreviations: TRAIL: TNF-related apoptosis-inducing ligand; TNFα: Tumor necrosis factor; IL-1: interleukin-1; NGF: nerve growth factor; Fas: Tumor necrosis factor receptor superfamily

member 6; TRAIL-R: TNF-related apoptosis-inducing receptor; TNF-R1: Tumor necrosis factor; IL-1R: interleukin-1 receptor; TrkA: transforming tyrosine kinase protein; FADD: Fas-associated death domain protein; TRADD: Tumor necrosis factor receptor type 1-associated DEATH domain protein; RIP1: receptor interacting protein 1; TRAF2: TNF receptor-associated factor 2; MyD88: myeloid differentiation primary response 88; IRAK: Interleukin-1 receptor-associated kinase 1; PI3K: Phosphoinositide 3-kinase; FLIP: FLICE-like inhibitory protein; IKK: IκB kinase; Akt/PKB: serine/threonine kinase Akt, also known as protein kinase B; PKA: protein kinase A; Cn: calcineurin; CASP8: caspase 8; Bid: BH3 interacting-domain death agonist; Bcl-2XL: B-cell lymphoma-extra-large; Apaf-1: Apoptotic protease activating factor 1; IkBa: nuclear factor of kappa light polypeptide gene enhancer in B-cells inhibitor, alpha; DFF45: (DNA fragmentation factor 45 kDa subunit; IAP: inhibitor of apoptosis; Bax: Bcl-2–associated X protein.

Figure 3. Examples of Ras-GTP complexes with Ras binding domains. A) Structures of Ras in the various complexes as well as representative structures of Ras alone (either in the GTP- or GDP-bound form) are superimposed at the center. GppNHp is used as the GTP analogue in most cases. All available complexes of human Ras with protein binding partners are depicted radiating from the center. Ras-GppNHp is depicted in shades of green and appears in the same color in the center and in the respective complexes. Ras-GDP is depicted in shades of pink, and apo Ras, which appears bound to SOS along with Ras-GTP, is in yellow. Ras binding domains of the various proteins that interact directly with Ras are shown in gray. Ras/GAP, PDB code 1WQ1; Ras/Nore1A, PDB code 3DDC; Ras/SOS, PDB code 1NVW; Ras/RalGDS, PDB code 1LFD; Ras/PLCε, PDB code 2C5L. Additional Ras structures, included to represent the range of conformational states observed for Ras, have the following PDB codes: GDP-bound form, 1LFD, 1LF5, 1ZVQ, 2QUZ, 2X1V, 3CON, 3LO5; GTP-bound form, 4EPR, 3K8Y, 2RGE, 4DLR. B) On the left is the complex between the Ras homologue Rap (cyan) and Raf-RBD (blue), where switch I of Rap has been mutated to the sequence in Ras, PDB code 1GUA. This is the best model currently available for the Ras-GTP/Raf complex. On the right is the KSR2/MEK1 complex, PDB code 2Y4I, that appears downstream from Ras in the Ras/Raf/MEK/ERK pathway and is usually not included in the KEGG style diagrams.

Figure 4. The catalytic domain of Ras in the GTP-bound form. A) Ribbon diagram with surface illustrating the two lobes of the molecule. The effector lobe, consisting of residues 1-86, is shown in green and contains all elements of the active site, with switch I, switch II and the P-loop labeled. The GTP analogue GppNHp is shown in orange with the stick representation. The allosteric lobe, consisting of residues 87-166, is in gray, with hot spots for protein/membrane interactions depicted with orange spheres and five-point starts. One of the sites is between helices 3 and 4 near the allosteric site occupied by calcium and acetate ions, and the other two are in the rear, situated between helices 4 and 5. Areas in these three pockets that are not identical in the isoforms of Ras are depicted in red. B) The Ras allosteric switch. The model in yellow represents the catalytically slow Ras conformation obtained in the absence of calcium

acetate (PDB code 2RGE). In this structure the switch II is disordered, with helix 3 sterically interfering with the C-terminal end of helix 2, an essential element in the ordered conformation of switch II. The model in green shows calcium and acetate bound in the allosteric site, promoting a shift in helix 3/loop 7 toward helix 4 and placement of R68 at the center of an H-bonding network that stabilizes the N-terminal end of switch II containing the catalytic residue Q61 (PDB code 3K8Y). R97 in helix 3 interacts directly with the acetate and plays a key role in promoting the shift associated with the allosteric switch mechanism. These and other residues involved are shown in stick and the crystallographic water molecules that are part of the allosteric switch network are shown as red spheres.

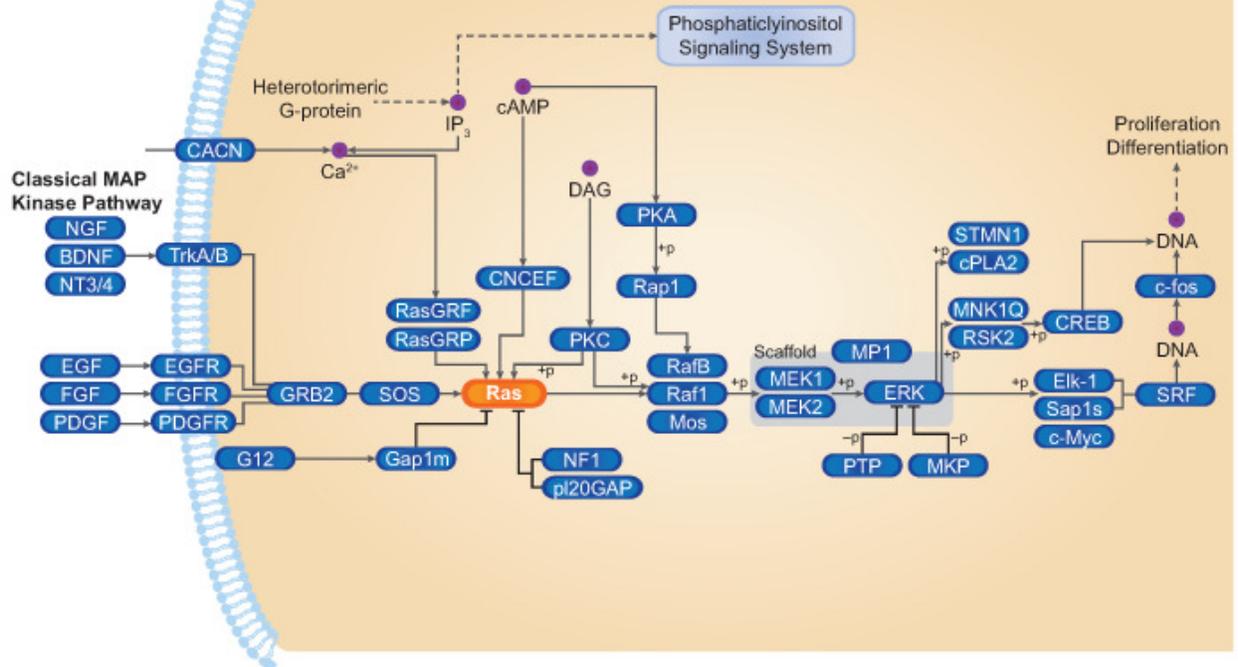

**Figure 1**

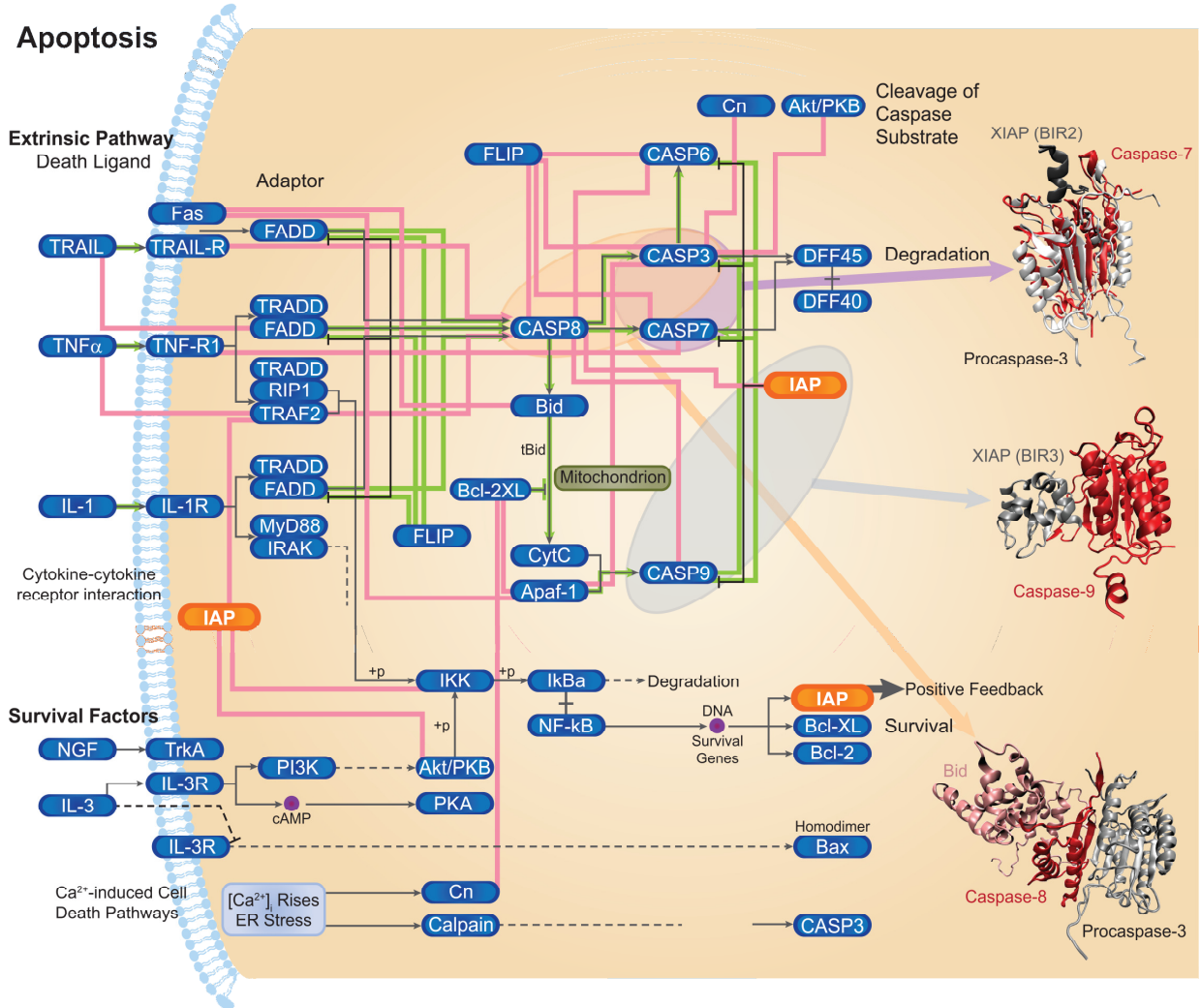

**Figure 2**

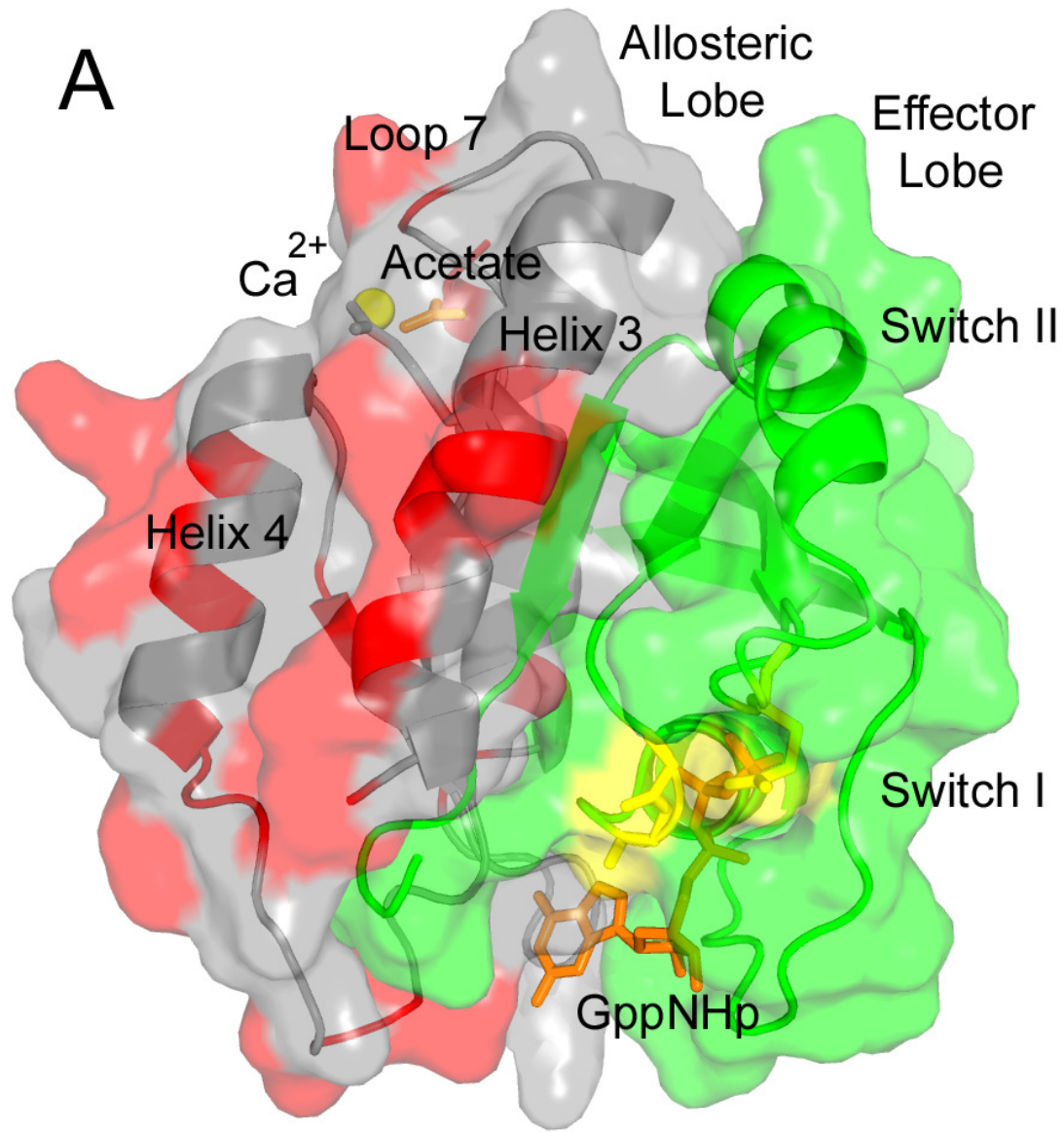

**Figure 3**

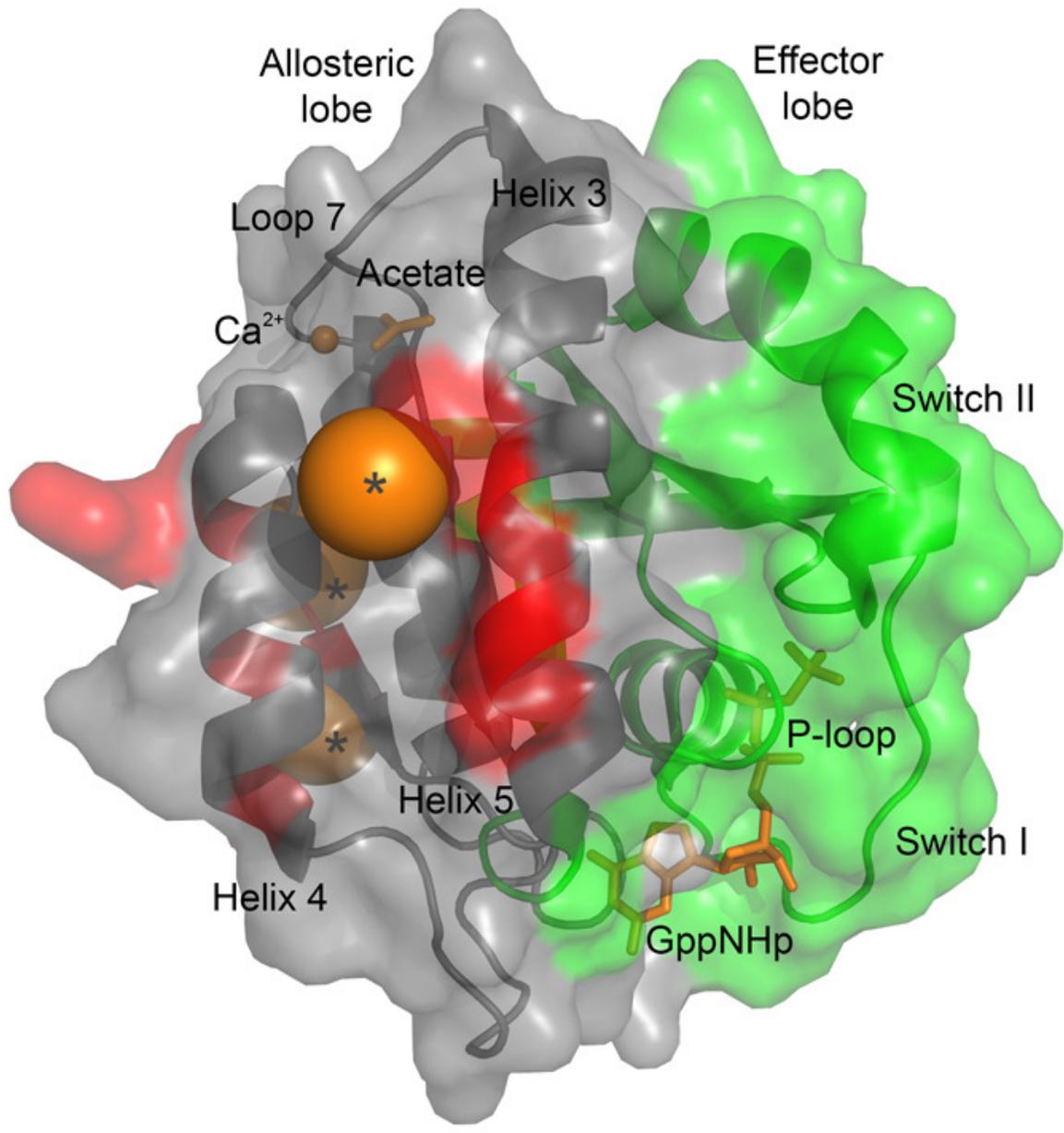

**Figure 4A**

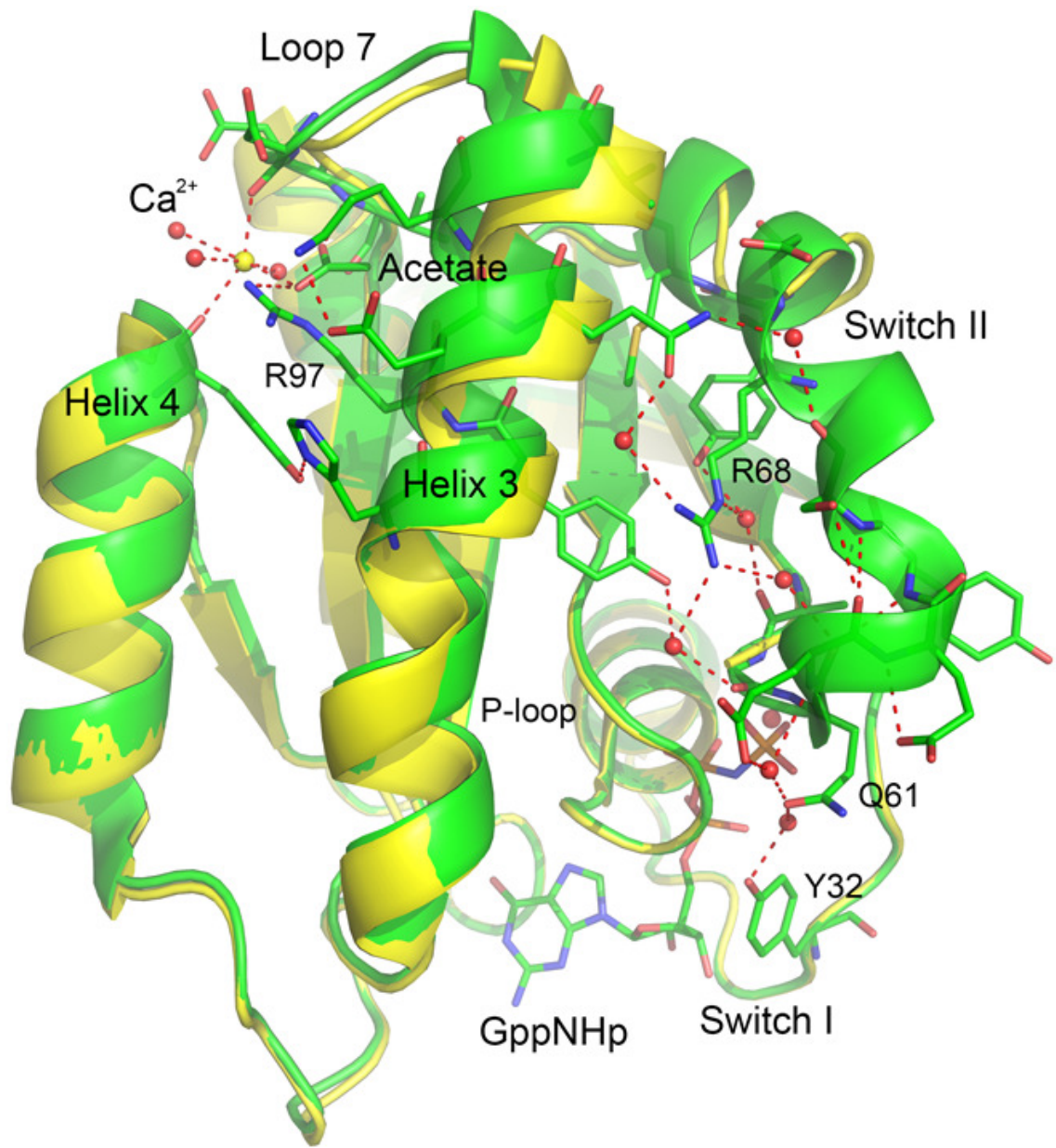

**Figure 4B**